\documentclass[prl,floatfix,showpacs,twocolumn,preprintnumbers,amsmath,amssymb,superscriptaddress]{revtex4}
\usepackage{graphicx,float,datetime}
\usepackage[utf8]{inputenc}
\usepackage{hyperref}

\newcommand{\udt}[3]{#1^{#2}_{\phantom{#2}#3}}

\newcommand{\dut}[3]{#1_{#2}^{\phantom{#2}#3}}

\newdateformat{mydate}{\THEDAY\hspace{3pt}\monthname[\THEMONTH] \THEYEAR}

\begin{document}

\begin{center}
\title{Reconstruction from scalar-tensor theory and the inhomogeneous equation of state in $f(T)$ Gravity}
\date{\mydate\today}
\author{Jackson Levi Said\footnote{jackson.said@um.edu.mt}}
\affiliation{Institute of Space Sciences and Astronomy, University of Malta, Msida, MSD 2080, Malta}
\affiliation{Department of Physics, University of Malta, Msida, MSD 2080, Malta}

\begin{abstract}
{
General relativity (GR) characterizes gravity as a geometric properly exhibited as curvature on spacetime. Teleprallelism describes gravity through torsional properties, and can reproduce GR at the level of equations. Similar to $f(R)$ gravity, on taking a generalization, $f(T)$ gravity can produce various modifications its gravitational mechanism. The resulting field equations are inherently distinct to $f(R)$ gravity in that they are second order. In the present work, $f(T)$ gravity is examined in the cosmological context with a number of solutions reconstructed by means of an auxiliary scalar field. To do this, various forms of the Hubble parameter are considered with an $f(T)$ lagrangian emerging for each instance. In addition, the inhomogeneous equation of state (EoS) is investigated with a particular Hubble parameter model used to show how this can be used to reconstruct the $f(T)$ lagrangian. Observationally, both the auxiliary scalar field or exotic terms in the FRW field equations give the same results, meaning that the variation in the Hubble parameter may be interpreted as the need to reformulate gravity in some way as is done in $f(T)$ gravity.
}
\end{abstract}


\maketitle

\end{center}

\section{I. Introduction}\label{sec:intro}
Modified gravity is one of the two direct approaches for reproducing the late time acceleration observed in the Universe \cite{Riess:1998cb,Perlmutter:1998np,Clifton:2011jh}. Additionally, there are other consistency problems that must eventually be tackled in the general relativity (GR) approach to gravity \cite{weinberg1972gravitation,Martin:2012bt}. The question then becomes what reformulation of gravity should be adopted, or whether we should take an extension of GR as our starting position. This is indeed a popular approach to the problem where many works consider an extension to the GR Einstein-Hilbert action in which the predominant attempt can be represented by the $f(R)$ gravity formulations \cite{Sotiriou:2008rp,DeFelice:2010aj}.\

One proposition that has gained interest in recent years is that of teleparallel gravity \cite{Cai:2015emx,Krssak:2015oua,Hayashi:1979qx,Paliathanasis:2016vsw}. The concept was initiated by Einstein himself shortly after his introduction of GR \cite{2005physics...3046U}. At first this was simply an equivalent reformulation of GR, called the {\it teleparallel equivalent of general relativity} (TEGR), which replaces the Einstein-Hilbert lagrangian with an analogue teleparallel torsion quantity, $T$ (known as the torsion scalar).\

As with the Einstein-Hilbert action and the $f(R)$ approach, the simplest generalization is to an arbitrary function of the torsion scalar, $f(T)$ \cite{Cai:2015emx}. Similarly, the TEGR formulation is equivalent to GR at the level of equations however its generalization is distinctly different from $f(R)$ in a number of important ways. The first being that the resulting field equations continue to be second order while those of $f(R)$ are fourth order. This has led to interesting insights both cosmologically and otherwise \cite{Paliathanasis:2016vsw,Krssak:2015oua,Cai:2015emx,Myrzakulov:2013wza,Rodrigues:2012qua}.\

Any alternative to GR should answer some of the open questions in GR. From the cosmological perspective, it would be interesting to put together a consistent theory that predicts the expansion behavior of both the early and late--time universe. In fact this is one of the primary motivations behind the $\Lambda$CDM modification to Einstein's GR formulation. While the initiation of the need for modifications to Einstein's GR comes from the observation in 1998 of the accelerating expansion of the universe \cite{Riess:1998cb,Perlmutter:1998np}, there is now overwhelming observational evidence for the need to modify the Einstein formulation \cite{Frieman:2008sn}.\

The goal of the present work is to produce known and new lagrangian models within the $f(T)$ gravity context using several Hubble parameter models \cite{Elizalde:2008yf}, using an auxiliary scalar field as a conduit to perform this reconstruction. This scalar field, essentially, takes on the role of dark energy, and would be the mediator through which these cosmological effects would take over.\

Unless stated otherwise, geometric units are used where $G=1=c$. Also Latin indices are used to refer to local inertial coordinates while Greek ones are used to refer to global coordinates. The paper is divided as follows, in section II, $f(T)$ gravity is introduced with some focus on the distinction between $f(R)$ and $f(T)$ gravity, and cosmological effects. In section III, the reconstruction work is presented along with the particular lagrangians that reproduce the Hubble parameters being considered. The inhomogeneous EoS is then derived in section IV where a particular example is given along with the general approach to determining $f(T)$ lagrangians. Finally the results are summarized and discussed in section V.

\section{II. $f(T)$ Gravity and cosmology}
GR and its variants are largely based on the metric tensor, $g_{\mu\nu}$, in terms of fundamental dynamical variables. This tensor acts as a potential quantity while the curvature is in turn represented through the Levi-Civita connection (which is torsion-free), $\Gamma^{\lambda}_{\mu\nu}$. In teleparallelism this connection is replaced by the Weitzenb\"{o}ck connection, $\hat{\Gamma}^{\lambda}_{\mu\nu}$. This new connection is curvature-free and is based on two fundamental dynamical variables, namely the tetrads (or vierbein) and the spin connection. The tetrads, $\udt{e}{a}{\mu}$, are four orthonormal vectors that relate inertial and global frames in that they build the metric up from the Minkowski metric by means of an application of this transformation. In real terms, they represent the physical observer and can be related to the metric by means of
\begin{equation}
g_{\mu\nu}=\eta_{ab}\udt{e}{a}{\mu}\udt{e}{b}{\nu}.
\end{equation}
where $\eta_{ab}=\text{diag}(1,-1,-1,-1)$. The tetrads obey the following inverse relations
\begin{equation}
\udt{e}{a}{\mu}\dut{e}{a}{\nu}=\delta^{\nu}_{\mu} \quad \udt{e}{a}{\mu}\dut{e}{b}{\mu}=\delta^{a}_{b}.
\end{equation}\

On the other hand, the spin connection, $\udt{\omega}{b}{a\mu}$, is not a tensor and is dependent on the particular system under consideration, i.e. it accounts for the coordinate system in a way retains the covariance of $f(T)$ gravity \cite{Krssak:2015oua}. Indeed in the literature there is a division between tetrads. There are those that produce a vanishing spin connection, $\udt{\omega}{b}{a\mu}=0$, called {\it pure} tetrads while in the {\it impure} tetrad frames the spin connection takes on some nonzero values.\

In this work we take the flat FRW metric
\begin{equation}
ds^2=dt^2-a^2(t)\Sigma_{i=1}^{3}dx_i^2,
\label{metric}
\end{equation}
where $x_i$ represent Cartesian coordinates. The natural choice of tetrad for this metric is
\begin{equation}
\udt{e}{a}{\mu}=\text{diag}(1,a(t),a(t),a(t)).
\label{frw_tetrad}
\end{equation}
Since this tetrad has vanishing spin connection components \cite{Krssak:2015oua}, no contributions needs to be included in what follows. In turn, the Weitzenb\"{o}ck connection takes on the form, $\hat{\Gamma}^{\lambda}_{\mu\nu}=\dut{e}{a}{\lambda}\partial_{\mu}\udt{e}{a}{\nu}$ \cite{Hayashi:1979qx}. This naturally leads to the torsion tensor
\begin{equation}
\udt{T}{\lambda}{\mu\nu}=\hat{\Gamma}^{\lambda}_{\mu\nu}-\hat{\Gamma}^{\lambda}_{\nu\mu}.
\end{equation}\

The difference between the Weitzenb\"{o}ck and the Levi-Civita connections is characterized by the contortion tensor
\begin{equation}
\udt{K}{\mu\nu}{a}=\frac{1}{2}\left(\dut{T}{a}{\mu\nu}+\udt{T}{\nu\mu}{a}-\udt{T}{\mu\nu}{a}\right).
\end{equation}
Lastly the superpotential tensor is introduced
\begin{equation}
\dut{S}{a}{\mu\nu}=\udt{K}{\mu\nu}{a}-\dut{e}{a}{\nu}\udt{T}{\alpha\mu}{\alpha}+\dut{e}{a}{\mu}\udt{T}{\alpha\nu}{\alpha},
\end{equation}
which is produced purely for mathematical convenience \cite{Cai:2015emx}.\

These tensors can be combined to form the torsion scalar, $T=\udt{T}{a}{\mu\nu}\dut{S}{a}{\mu\nu}$, which is the lagrangian for TEGR.

It is at this point that the distinction between curvature and this torsion-based theory of gravity can be made clearer. In GR we adopt the Ricci scalar, $R$, as the gravitational lagrangian \cite{misner1973gravitation} whereas in the TEGR case the torsion scalar is taken. The difference between the two quantities obviously lies in a boundary term since they produce the same theory at the level of equations \cite{Cai:2015emx}. The distinction can be quantified through \cite{Bahamonde:2015zma}
\begin{equation}
R(e)=-T+B,
\end{equation}
where $B=\frac{2}{e}\partial^{\mu}\left(e\udt{T}{\lambda}{\lambda\mu}\right)=2\nabla^{\mu}\udt{T}{\lambda}{\lambda\mu}$ is the boundary term. Therefore, taking the lagrangian $-T+B$ will exactly reproduce the Ricci scalar. However, this similarity is lost once the theory is generalized to an arbitrary function thereof, that is, unless the function is $f(R)=f(-T+B)$. In all other cases, $f(R)$ and $f(T)$ will not coincide in any meaningful way. Moreover, the ensuing field equations are unique in that, out of the three possible quantities involved, namely $R$, $T$, and $B$, $f(T)$ is the only lagrangian that produces second order field equations \cite{Bahamonde:2015zma,Bahamonde:2016cul}.\

Now, using the tetrad in Eq.(\ref{frw_tetrad}), the torsion scalar turns out to be  \cite{Krssak:2015oua}
\begin{equation}
T=-12\frac{\dot{a}^2(t)}{a^2(t)}\equiv -12H^2.
\label{torsion_scalar}
\end{equation}
Generalizing the action to an arbitrary function of the torsion scalar, $f(T)$, results in
\begin{equation}
S=\frac{1}{4\kappa}\displaystyle\int d^4 x e f(T),
\label{torsion_action}
\end{equation}
where $\kappa=4\pi G$ and $e=\det{\udt{e}{a}{\mu}}$. On taking a variation with respect to the tetrad the following field equations emerge \cite{Cai:2015emx}
\begin{align}
\dut{E}{a}{\mu}&\equiv  e^{-1} f_{T} \partial_{\nu}\left(e \dut{S}{a}{\mu\nu}\right)+f_{TT} \dut{S}{a}{\mu\nu} \partial_{\nu} T\nonumber\\
&-f_{T} \udt{T}{b}{\nu a}\dut{S}{b}{\nu\mu}+\frac{1}{4}f(T)\dut{e}{a}{\mu}=\kappa \dut{\Theta}{a}{\mu},
\end{align}
where $\dut{\Theta}{a}{\mu}\equiv \frac{1}{e}\frac{\delta \mathcal{L}_m}{\delta \udt{e}{a}{\mu}}$, $f_{T}$ and $f_{TT}$ denote the first and second derivatives of $f(T)$ with respect to $T$, and $\mathcal{L}_m$ is the matter lagrangian.\

As usual, for the cosmological case a perfect fluid is assumed, i.e
\begin{equation}
\Theta_{\mu\nu}=\left(p_m(t)+\rho_m(t)\right)u_{\mu}u_{\nu}-p_m g_{\mu\nu},
\end{equation}
where $p_m(t)$ and $\rho_m(t)$ are the matter pressure and energy density respectively, and $u_{\mu}$ is the fluid four velocity $u_{\mu}=(1,0,0,0)$.\

The resulting Friedmann equations turn out to be \cite{Krssak:2015oua}
\begin{equation}
\frac{f}{4}+6f_{T} H^2=\kappa \rho_m,
\label{field_1}
\end{equation}
and
\begin{equation}
\frac{f}{4}+6f_T H^2+2f_{T} \dot{H}-48 f_{TT} H^2 \dot{H}=-\kappa p_m,
\label{field_2}
\end{equation}
where dots denote derivatives with respect to cosmic time.\

With these governing equations in hand, an auxiliary scalar field can now be introduced to reconstruct the $f(T)$ lagrangian.

\section{III. Reconstruction in $f(T)$ gravity using scalar-tensor theory}
In general it is difficult to determine specific cosmologies directly from the modified Friedmann equations in Eqs.(\ref{field_1},\ref{field_2}) due to the general nature of these relations. For this reason a technique first developed for $f(R)$ gravity in Ref.\cite{SaezGomez:2008uj} is investigated in the $f(T)$ setting. Following Refs.(\cite{SaezGomez:2008uj, Nojiri:2003ft, Nojiri:2007cq}) an auxiliary scalar field, $\phi$, without a kinetic term is explored. This approach has shown to be very effective in determining solutions to otherwise insoluble equations.\

\noindent In this context consider the action
\begin{equation}
S=\displaystyle\int d^4 x e\left(P(\phi)T+Q(\phi)\right),
\label{scalar_action}
\end{equation}
where $P(\phi)$ and $Q(\phi)$ are arbitrary functions.\

Taking a variation with respect to the tetrads, $\udt{e}{a}{\mu}$, gives
\begin{equation}
\dut{h}{a}{\mu}\left(P(\phi)T+Q(\phi)\right)-4P(\phi)\udt{T}{b}{\nu a}\dut{S}{b}{\nu\mu}=\kappa \dut{\Theta}{a}{\mu}.
\label{scalar_field_eqns}
\end{equation}
The benefit of taking this scalar field form is that a second variation can be taken, one with respect to the scalar field itself. This results in the simple relation
\begin{equation}
P'(\phi)T+Q'(\phi)=0,
\end{equation}
where primes denote derivatives with respect to the scalar field, $\phi$. This second relation can in turn be used to determine a relation between the scalar field and the torsion scalar, $\phi=\phi(T)$. Replacing the result in the action in Eq.(\ref{scalar_action}) gives a way to resolve the lagrangian in the original action in Eq.(\ref{torsion_action})
\begin{equation}
f(T)=P(\phi(T))T+Q(\phi(T)).
\label{phi_T_relation}
\end{equation}\

The scalar field governing relations in Eq.(\ref{scalar_field_eqns}) can be taken for any setting, in particular we take them for the metric in Eq.(\ref{metric}) which results
\begin{align}
12H^2 P(\phi)+Q(\phi)&=\kappa \rho_m,\nonumber\\
Q(\phi)-10H^2 P(\phi)&=-\kappa p_m,
\end{align}
which are the Friedmann equations for the scalar field.\

The scalar field can easily be taken to be the coordinate time, $\phi=t$, and setting the EoS to be $p_m=\omega_m \rho_m$. The stress-energy tensor has not been altered in any significant way so the same conservation equation naturally follows, namely $\dot{\rho}_m+3H(1+\omega_m)\rho_m=0$. Solving this relation naturally leads to the cosmic time dependence of the energy density
\begin{equation}
\rho_m=\rho_{m,0} \text{Exp}\Big[-3(1+\omega_m)\displaystyle\int dt H(t)\Big].
\end{equation}\

The solution for $\rho_m$ is important because with this in hand, the Friedmann equations for the scalar field can be used to determine the scalar field action components as a function of the Hubble parameter. However, in this case the Hubble parameter will be a function of the scalar field, $H=g(\phi)$, which is where the relationship between the torsion scalar and the scalar field comes in. Using Eqs.(\ref{scalar_field_eqns}) and taking the coupling parameters as $\kappa=1=\rho_{m,0}$ for convenience, the scalar field functions can be determined. Since the theory is second order, the scalar field turns out to be soluble in terms of the Hubble parameter. After solving the coupled differential equations
\begin{equation}
P(\phi) g^2(\phi) \text{Exp}\Big[\frac{3\left(1+\omega_m\right)}{2} g^2(\phi) \Big]=k,
\label{P_equation}
\end{equation}
where $k$ is some constant, and
\begin{align}
Q(\phi)&=10 g^2(\phi)P(\phi)\nonumber\\
&-\omega_m \text{Exp}\Big[-3(1+\omega_m)\displaystyle\int d\phi g(\phi)\Big].
\label{Q_equation}
\end{align}\

For any given scalar field functional dependence, $P=P(\phi)$, Eq.(\ref{P_equation}) returns a Lambert-W function so this is not considered. Instead various Hubble parameter profiles are taken in order to investigate the reconstruction advantages of this approach. In the following these Hubble parameter model are introduced along with the associated motivation, the Lagrangian functions are then worked out.

\subsubsection{Model 1}
First a two term Hubble parameter which takes different forms at early and late times is considered. Take \cite{Elizalde:2008yf,SaezGomez:2008uj}
\begin{equation}
g(\phi)=\frac{H_1}{\phi^2}+\frac{H_0}{t_s-\phi},
\end{equation}
where $t_s$ is an arbitrary characteristic time that represents the time at which the big rip would occur. At early times this Hubble parameter leads to a period of accelerated expansion similar to inflation which is followed directly by a decelerated period of expansion. As $t\rightarrow t_s$, the expansion of the universe becomes super-accelerated leading to the big rip. This is one straightforward way to reconcile the early- and late-times behavior of the universe.\

For early times, $t<<t_s$, it straightforwardly follows that
\begin{align}
H(t)&\sim\frac{H_1}{t^2}\\
\frac{\ddot{a}}{a}&\sim \frac{H_1}{t^2}\left(\frac{H_1}{t^2}-\frac{2}{t}+\frac{2H_0}{t_s-t}\right)>0.
\end{align}
Similar to the concept of inflation, for a time $t$ close to zero, the acceleration parameter is positive, $\frac{\ddot{a}}{a}>0$. The model then naturally enters a period of deceleration \cite{Elizalde:2008yf} while the condition $t<<t_s$ is retained.\

The scalar field functions can then be determined which will turnout to fix the $f(T)$ lagrangian. Firstly, by Eq.(\ref{P_equation}) it is found that
\begin{equation}
P(\phi)=\frac{k\,\phi^4}{H_1^2 \text{Exp}[\frac{3(1+\omega_m)}{2}\left(\frac{H_1}{\phi^2}\right)^2]},
\end{equation}
and by Eq.(\ref{Q_equation}) it follows that
\begin{align}
Q(\phi)&=\frac{10\,k}{\text{Exp}\Big[\frac{3(1+\omega_m)}{2}\left(\frac{H_1}{\phi^2}\right)^2\Big]}\nonumber\\
&-\omega_m\text{Exp}\Big[\frac{3(1+\omega_m)H_1}{\phi^2}\Big].
\end{align}
By using Eq.(\ref{torsion_scalar}) the scalar field can be expressed in terms of the torsion scalar, $\phi^2=-12H_1/T$. In turn, this leads to a functional form for the original $f=f(T)$ lagrangian through the scalar field in the lagrangian of Eq.(\ref{scalar_action}), giving
\begin{equation}
f(T)\sim-\omega_m \text{Exp}\Big[-\frac{1+\omega_m}{4}\,T\Big].
\end{equation}
This represents the active part of the lagrangian that would be needed to account for the early time segment of the universe in $f(T)$ gravity.\

Now for very late times, approaching the big rip, i.e $t\rightarrow t_s$ resulting in Hubble and acceleration parameters
\begin{align}
H(t)&\sim\frac{H_0}{t_s-\phi}\nonumber\\
\frac{\ddot{a}}{a}&\sim\frac{H_0}{(t_s-\phi)^2}(1+H_0).
\end{align}
As the big rip time is approached the universe enters a phase of very rapid expansion again \cite{Elizalde:2008yf}. As shown above, the scalar field functions can be determined through Eq.(\ref{P_equation}) and Eq.(\ref{Q_equation})
\begin{equation}
P(\phi)=\frac{k\,(t_s-\phi)^2}{H_0^2\text{Exp}\Big[\frac{3(1+\omega_m)H_0^2}{2(t_s-\phi)^2}\Big]},
\end{equation}
and analogously
\begin{align}
Q(\phi)&=\frac{10\,k}{\text{Exp}\Big[\frac{3(1+\omega_m)H_0^2}{2(t_s-\phi)^2}\Big]}\nonumber\\
&-\omega_m\left(t_s-\phi\right)^{3H_0(1+\omega_m)}.
\end{align}
Inverting the torsion scalar in Eq.(\ref{torsion_scalar}) results in a reconstructed function for the torsion lagrangian
\begin{equation}
f(T)\sim-10k\,\text{Exp}\Big[\frac{1}{6}(1+\omega_m)T\Big].
\end{equation}
In this case the additional component of the lagrangian is again exponential however in this case it is positive since it plays a role at late times.\

The intermediary phase of the universe would still be accounted for by the TEGR component. The end result is a lagrangian for $f(T)$ gravity given by $f(T)= T - a_0\text{Exp}[-A_0 T] - b_0\text{Exp}[A_1 T]$, where the $a_i$ and $A_i$ are positive constants. The resulting theory would satisfy the cosmological requirements. However more work would have to be done to understand the behavior of the theory on smaller scales such as for galaxies and the solar system. One way to get a more constrained lagrangian from this approximate form would be to compare supernova data and reduce the arbitrariness of the free parameters.

\subsubsection{Model 2}
In model 1 the very late-time rapid acceleration takes on a phantom nature. Another approach is to consider \cite{Elizalde:2008yf}
\begin{equation}
g(\phi)=H_0+\frac{H_1}{\phi^n},
\end{equation}
where $H_0$ and $H_1>0$ are constants, and $n$ is a positive integer. The $n>1$ region is considered since it gives three well defined regions corresponding to an early acceleration phase (interpreted as inflation), then deceleration, and finally late-time acceleration. The corresponding acceleration parameter turns out to be
\begin{equation}
\frac{\ddot{a}}{a}=\dot{H}+H^2=-\frac{nH_1}{\phi^{n+1}}+\left(H_0+\frac{H_1}{\phi^n}\right)^2.
\end{equation}
For the model under consideration we do not need to take limits to determine the $f(T)$ lagrangian, so the scalar functions can straightforwardly be determined giving
\begin{align}
P(\phi)&=\frac{k}{\left(H_0+\frac{H_1}{\phi^n}\right)^2\text{Exp}\Big[\frac{3(1+\omega_m)}{2}\left(H_0+\frac{H_1}{\phi^n}\right)^2\Big]},\nonumber\\
&
\end{align}
and
\begin{align}
Q&(\phi)=10k\,\text{Exp}\Big[-\frac{3(1+\omega_m)}{2}\left(H_0+\frac{H_1}{\phi^n}\right)^2\Big]\nonumber\\
&-\omega_m\text{Exp}\Big[-3(1+\omega_m)\left(H_0\phi-\frac{H_1}{(n-1)\phi^{n-1}}\right)\Big].\nonumber\\
&
\end{align}
Inverting the torsion scalar relation in Eq.(\ref{torsion_scalar}) results in a scalar field
\begin{equation}
\phi=\left(\frac{H_1}{\sqrt{-\frac{T}{12}}-H_0}\right)^{1/n}.
\end{equation}
The arbitrary function $f(T)$ can then be determined in general
\begin{align}
f(T)&=P(\phi)T+Q(\phi)\nonumber\\
&=-2k\,\text{Exp}\Big[\frac{1+\omega_m}{8}T\Big]\nonumber\\
&-\omega_m\,\text{Exp}\Big[\frac{3(1+\omega_m)}{1-n}\left(n H_0-\sqrt{-T/12}\right)\nonumber\\
&\left(\frac{H_1}{\sqrt{-T/12}-H_0}\right)^{1/n}\Big].
\end{align}\

As in the first model the TEGR term is still included to govern the intermediate stage of evolution of the universe, resulting in the general lagrangian form $f(T)\sim T+b_0 \text{Exp}[b_1 T]+B_0 \text{Exp}[B_1(B_2+\sqrt{-T})\left(\frac{1}{\sqrt{-T}+B_3}\right)^N]$. The square root term actually emerges naturally in other models \cite{Rodrigues:2012qua,Myrzakulov:2013wza,Paliathanasis:2016vsw} however when it appears on its own it does not contribute to the eventual Friedmann equations whereas in this case it is implicit in another function. In the current context it is interesting for it to turn out to form part of the lagrangian solution. To a much lesser extent this is not dissimilar to the Gauss-Bonnet generalizations \cite{delaCruzDombriz:2011wn}.

\subsubsection{Model 3}
Lastly, the following model is considered \cite{Elizalde:2008yf}
\begin{equation}
g(\phi)=\frac{H_i+H_l c e^{2\alpha\phi}}{1+c e^{2\alpha\phi}},
\end{equation}
where $H_i$, $H_l$, $c$ and $\alpha$ are positive constants. In model 2 additional scalar fields have to be introduced to suppress issues with inflation whereas in the current context a clear link with observation is attainable.\

For the early and late universe the hubble parameter tends to the constants $H_i$ and $H_l$ respectively. In this way, $H_i$ would drive inflation and $H_l$ would take on the small cosmological constant for late times.\

The corresponding acceleration parameter turns out to be
\begin{align}
&\frac{\ddot{a}}{a}=\nonumber\\
&\frac{2\alpha c e^{2\alpha\phi}(H_l-H_i)+(H_i+cH_l e^{2\alpha\phi})^2}{(1+c e^{2\alpha\phi})^2}.\nonumber\\
&
\end{align}\

Putting this into the scalar field functions in the lagrangian in Eq.(\ref{P_equation}) and Eq.(\ref{Q_equation}) results in
\begin{align}
P(\phi)&=\nonumber\\
&\frac{k\left(1+c e^{2\alpha\phi}\right)^2}{(H_i+H_l c e^{2\alpha\phi})^2\text{Exp}\Big[\frac{3(1+\omega_m)}{2}\left(\frac{H_i+H_l c e^{2\alpha\phi}}{1+c e^{2\alpha\phi}}\right)^2\Big]},\nonumber\\
&
\end{align}
and
\begin{align}
Q(\phi)&=\nonumber\\
&\frac{10k}{\text{Exp}\Big[\frac{3(1+\omega_m)}{2}\left(\frac{H_i+H_l c e^{2\alpha\phi}}{1+c e^{2\alpha\phi}}\right)^2\Big]}\nonumber\\
&-\omega_m \text{Exp}\Big[-3 H_i \phi \left(1+\omega_m\right)\Big]\left(1+c e^{2\alpha \phi}\right)^B,
\end{align}
where $B=-\frac{3(1+\omega_m)}{2\alpha}(H_l-H_i)$. As in the previous case, the relationship between the scalar field and the torsion scalar follows straightforwardly
\begin{equation}
\phi=\frac{1}{2\alpha}\ln\left(\frac{1}{c}\frac{H_i-1}{\sqrt{-\frac{T}{12}}-H_l}\right).
\end{equation}
Through the mechanics of the scalar field, $\phi$, Eq.(\ref{phi_T_relation}) is utilized to produce the teleparallel lagrangian
\begin{align}
f&(T)=-2k\text{Exp}\left[\frac{(1+\omega_m)T}{8}\right]\nonumber\\
&-\omega_m\left(1+\frac{H_i-1}{\sqrt{-\frac{T}{12}}-H_l}\right)\left(\frac{1}{c}\frac{H_i-1}{\sqrt{-\frac{T}{12}}-H_l}\right)^B.
\end{align}

Incorporating TEGR again gives a relatively simple lagrangian that contains different lagrangian elements for the different epochs of the cosmological history. In this instance this gives the model $f(T)\sim T+d_0 \text{Exp}[d_1 T]+D_0\left(1+\frac{H_i-1}{\sqrt{-\frac{T}{12}}-H_l}\right)\left(\frac{H_i-1}{\sqrt{-\frac{T}{12}}-H_l}\right)^D_1$ which has two effective Hubble parameters for early and late times. This last model can work as a bridge between the first two model since it contains both terms in $\sqrt{-T}$ and also the exponential term in $T$. Altogether a wide variety of potential lagrangian terms emerge from considering this collection of Hubble parameter ansatz.

\section{IV. The inhomogeneous equation of state}
In the previous section different Hubble functions are used to reconstruct the gravitational lagrangian using a scale factor approach as the vehicle for reconstruction. In this section the additional terms that $f(T)$ offers are interpreted as an extra dark fluid. The resulting EoS of the dark fluid depends on the Hubble parameter and its derivatives, in an inhomogeneous way.\

Taking the modified Friedmann equations from Eqs.(\ref{field_1},\ref{field_2})
\begin{align}
3H^2&=\frac{1}{f_T}\left(-\frac{f}{8}\right),\nonumber\\
-3H^2-2\dot{H}&=\frac{1}{f_T}\left(\frac{f}{2}-24 f_{TT}H^2\dot{H}\right),
\end{align}
where no other matter contributions are admitted. This would represent a cosmology dominated by modified gravity terms. On comparison with the standard Friedmann equations from GR, the energy and pressure densities contributions can be easily identified as
\begin{align}
\rho&=\frac{1}{\kappa}\frac{1}{f_T}\left(-\frac{f}{8}\right)\nonumber\\
p&=\frac{1}{\kappa}\frac{1}{f_T}\left(\frac{f}{8}-24f_{TT} H^2 \dot{H}\right),
\end{align}
where a perfect fluid context is being assumed. That is, the modified terms can be interpreted as forming part of a cosmic perfect fluid that takes on the role of dark energy. Following this line of thought a dark fluid EoS naturally emerges as
\begin{equation}
\omega=\frac{p}{\rho}=-1+\frac{192 f_{TT} H^2 \dot{H}}{f},
\end{equation}
which can be represented as
\begin{equation}
p=-\rho-\frac{24f_{TT}H^2\dot{H}}{f_T},
\end{equation}
where the second term is dependent only on the Hubble parameter ($\kappa=1$). This is the inhomogeneous EoS for the dark fluid which has been investigated in a number of works \cite{Brevik:2007jt,SaezGomez:2008uj,Nojiri:2006zh,Capozziello:2005pa,Nojiri:2005sr}. Following Refs.\cite{SaezGomez:2008uj,SaezGomez:2008qe}, this equation can be generalized to
\begin{equation}
p=-\rho+g\left(H,\dot{H},\ddot{H},...\right),
\label{inhomogeneous_eq}
\end{equation}
where $g\left(H,\dot{H},\ddot{H},...\right)=-\frac{24 f_{TT} H^2\dot{H}}{f_T}$. On combining the Friedmann equations and using the inhomogeneous EoS gives the following differential equation in $f(T)$
\begin{equation}
\dot{H}+\frac{\kappa}{2}g\left(H,\dot{H},\ddot{H},...\right)=0,
\label{inhomo_ode}
\end{equation}
where $\kappa$ is taken as unity for convenience in what follows. This equation can be solved only once a Hubble parameter is assigned since it is essentially a linear first order differential equation in $f(T)$.\

As an example the following Hubble parameter is considered \cite{SaezGomez:2008qe}
\begin{equation}
H(t)=H_0 t + \frac{H_1}{t},
\end{equation}
where $H_0$ and $H_1$ are positive constants. The Hubble model gives an early decelerating period followed up an accelerating period, which are studied in in turn as limiting cases. Another reason why this is done is because the relationship between the torsion scalar and cosmic time is non-invertible in the general case. This is needed to eliminate any appearances of cosmic time as $t=t(T)$ in the eventual $f(T)$ lagrangian.\

For the initial decelerating period, $H(t)\sim H_1/t$, giving a time dependence
\begin{equation}
t=H_1\sqrt{\frac{12}{-T}},
\end{equation}
and solving Eq.(\ref{inhomo_ode}) gives
\begin{equation}
f(T)=-144 H_1^4 \lambda \sqrt{\pi} \text{Erfi}\left[\frac{T^2}{288 H_1^4}\right],
\end{equation}
where $\lambda$ is an integration constant. It is not uncommon for $f(T)$ gravity to produce lagrangians that cannot be put in closed form \cite{Odintsov:2015uca,Wei:2015oua,Cai:2015emx}. Thus by using Eq.(\ref{inhomogeneous_eq}), the inhomogenous EoS equation that turns out to reproduce this kind of behavior is given by
\begin{equation}
p\sim -\rho- \frac{H^2\dot{H}}{6 H_1^4}\,T.
\end{equation}\

On the other hand, for late-times the Hubble parameter takes the form $H(t)\sim H_0 t$, and combining with Eq.(\ref{torsion_scalar}) results in the time dependence
\begin{equation}
t=\sqrt{-\frac{T}{12 H_0^2}}.
\end{equation}
Again solving the differential equation Eq.(\ref{inhomo_ode}) turns out to give a lagrangian
\begin{equation}
f(T)=\tilde{\lambda}\,\text{ln}T,
\label{second_f_inhomo}
\end{equation}
where $\tilde{\lambda}$ is an integration constant. This general type of lagrangian has shown some promise in cosmological settings \cite{Odintsov:2015uca,Cai:2015emx}. Taking derivatives gives the general inhomogeneous EoS equation as
\begin{equation}
p=-\rho+\frac{24 H^2\dot{H}}{T}.
\end{equation}
This indeed reproduces the desired late time behavior.

\section{V. Discussion and conclusion}
In this work a number of different solutions of $f(T)$ theory have emerged from investigating the cosmological scenario within the context of both early and late time epochs. This was achieved through the vehicle of using an auxiliary scalar similar to the  approach \cite{SaezGomez:2008uj}. This method works very well for generalized theories because it gives an second way of solving the Friedmann equations.\

Secondly, the inhomogeneous EoS was also derived using a particular Hubble parameter, with both early and late time reconstructions. Together these two approaches offer two methods of using an auxiliary scalar field to reconstruct the $f(T)$ lagrangian. Moreover, both the auxiliary scalar field and exotic modified gravity terms interpreted as a perfect fluid EoS result in the same state parameter behavior.\

$f(T)$ theories may provide a suitable description of the expansion history of the universe without the need to include exotic components to the stress-energy part. The models would need to be constrained at lower scales, such as in the galactic and solar system regime. Following this the models should next be compared with cosmological data such as supernovae luminosity distance and the CMB profile, similar to the initiative in $f(R)$ theory, however this is out of the scope of this paper.\

Beyond the models considered for the first method another two were attempted, namely $g(\phi)=\phi^{\alpha}$ and $g(\phi)=h_0^2 \left(\frac{1}{t_0^2-\phi^2}+\frac{1}{t_1^2+\phi^2}\right)$ however in either case the problem becomes intractable very fast. Other approaches may include perturbative analyses.\

For the models that do emerge, firstly they are predominantly exponential functions but some are not. In a follow-up study, it would be interesting to understand better the transition between the different epochs in terms of the lagrangian terms presented here. This would naturally involve an analysis of the stability of each model.\

\end{document}